\magnification=1200 
\def\al{\alpha}
\def\th{\theta}
 
\def\Dl{\Delta}

\def\omo{\Omega}

\rightline{$\,$}
\medskip 
\centerline{\bf Averaging Inhomogeneous Universes: Volume, Angle, Line of 
Sight} 
\bigskip 
\centerline{Eric V.~Linder} 
\centerline{University of Massachusetts, Department of Physics and Astronomy, 
Amherst, MA 01003}
\bigskip 
\centerline{ABSTRACT}
\medskip 
Cosmologies that match in a volume averaged sense 
need not generally have the same light propagation behaviors.  In particular 
a universe with inhomogeneity may not demonstrate the 
Friedmann-Robertson-Walker distance-redshift relation even after volume 
averaging to FRW spacetime.  Even the 
Dyer-Roeder prescription for incorporating inhomogeneity within a universe 
equivalent to FRW in an angle averaged sense does not guarantee FRW 
behavior in general.  To legitimately use the FRW distance-redshift 
relation to interpret observations, the physical conditions must 
match in a line of sight sense (defined herein: most observations do), 
since light 
probes the mass distribution or geometry on all length scales. 
\bigskip 

\centerline{\bf 1. Introduction}
\medskip 
Since many cosmological 
quantities depend on the distance-redshift relation it is of great 
observational relevance to ensure that the Friedmann-Robertson-Walker (FRW) 
expression is the correct one to use.  For our universe, containing local 
inhomogeneities, 
it is generally assumed that implementing some sort of large scale averaging 
procedure causes the behavior of light propagation to revert 
to FRW.  In the absence of 
a unique and rigorous mathematical prescription for averaging 
(Buchert 1997; Buchert \& Ehlers 1997), due to 
the nonlinear complexities of general relativity, one tends to rely on 
simplified models and reasonability arguments. 

One commonly adopted model is that of Dyer and Roeder (1972, 1973: DR), which 
restricts the inhomogeneities to randomly distributed compact clumps and 
compensates for the 
higher density within the clumps with a lower density of smoothly distributed 
matter -- so the overall density in a volume containing on average many 
clumps is on average the same as the equivalent smooth FRW model.  
Nevertheless, the angular 
diameter distance-redshift relations differ in the two cases and 
can cause, for example, ``shrinking'' of angular scales (Linder 1988b) 
in an inhomogeneous universe. 
A typical 
line of sight will avoid the clumps and hence the obvious gravitational 
lensing effects, making it difficult for the observer to discern whether 
the universe is truly smooth FRW or actually clumpy DR, and hence which 
distance relation is appropriate. 

While simplistic, this model allows us to calculate easily 
the distance-redshift relation for such an inhomogeneous universe, and  
evaluate a criterion for what type of averaging can restore the FRW result 
(Linder 1998); increasing the angular size of the observation 
beam encompasses more and more clumps and causes a transition toward the FRW 
behavior.  (For numerical investigation of distances in clumpy 
universes see Fukushige \& Makino 1994; Watanabe \& Tomita 1990).  

One can ask whether general inhomogeneous models that in some sense are 
``close'' to FRW can be realistically averaged to transform their 
inhomogeneous distance-redshift relations  into the FRW one in some limit.  
The interesting results of Mustapha, Bassett, Hellaby, \& Ellis (1998: MBHE) 
show that even a full sky average, i.e.~isotropization of an 
inhomogeneous universe, may not give the FRW relation generally. 

MBHE work within a spherically symmetric, 
Tolman-Bondi model and find that the FRW results are also not recovered even 
at source depths much greater than the inhomogeneity depth.  Because the 
calculations are already in an isotropic framework one cannot appeal 
to angular averaging to remove any discrepancies from FRW.  And 
volume averaging, over a large region of space that has overall properties 
equal to its FRW analog, is found insufficient to yield equal light 
propagation behaviors.  

These problems with averaging arise because the light is affected by the 
spacetime, 
the gravitational field of the matter along its path, on all scales and not 
just by some coarse grained quantity.  [Dynamics, on the other hand, 
such as the expansion of a region or growth of density perturbations, 
could be expected to be reasonably handled by a properly volume smoothed 
description (see Bildhauer \& Futamase 1991; Buchert \& Ehlers 1997).]  

This paper examines what averaging procedures can be applied to various 
inhomogeneous universes in order to attain the FRW distance-redshift behavior. 
Section 2 first investigates some formal concerns regarding large scale 
dynamics and flux conservation conditions, then section 3 considers volume 
(\S3.1) and angle (\S3.2) averaging, and introduces line of sight averaging 
(\S3.3), 
and section 4 summarizes the divergence/transition conditions to FRW.  

\bigskip 
\centerline{\bf 2. Dynamical Deviation and Flux Conservation}
\medskip 
If we want the deviation of the distance-redshift behavior of the 
inhomogeneous model from FRW to be large enough 
to be observationally significant, we must check whether 
such inhomogeneity will affect the dynamics of the volume we are concerned 
with.  The DR model has an equation for the light beam area, or equivalently 
angular diameter distance $r$, of 
$$
r''/r=-(1/2) R_{ab}k^ak^b=-(3/2)(1+z)^2H^2(z)\al\,\omo(z),\eqno(1)
$$ 
with respect to affine parameter $\lambda$, or 
$$
\ddot r+[3+q(z)](1+z)^{-1}\dot r+(3/2)(1+z)^{-2}\al\,\omo(z)\,r=0\,,\eqno(2)
$$ 
with respect to redshift $z$.  Here $R_{ab}$ is the Ricci tensor, $k^a$ the 
photon four momentum, $H(z)$ the Hubble parameter, $q(z)$ the deceleration 
parameter, and $\omo(z)$ the density 
parameter.  The smoothness parameter $\al$ is the 
ratio of smoothly distributed mass density to the total 
mass density.  (The forms of eqs.~1-2 do not change for a universe with 
components other than dust -- see Linder 1988a.) 

However, if the background FRW dynamics is affected, we must 
abandon the DR use of the global values for the Hubble, deceleration, and 
large scale density parameters and use instead the local values that the light 
experiences along its path.  That is, while in the DR model the difference of 
the local density from the FRW background was sufficiently represented in the 
effective density $\alpha\Omega$, now the density inhomogeneity also enters 
significantly into (the local) $H$ and $q$.  

At first glance, it is not as bad as one might think.  The beam equation 
in terms of the Ricci tensor (first part of equation 1) still holds (we are 
treating the case of the inhomogeneous universe being sufficiently close to 
FRW that shear can still be neglected --a circular beam remains circular -- 
thus allowing a single quantity $r$ to describe the angular diameter 
distance).  Likewise, its conversion to being 
written in terms of the energy-momentum tensor: 
$$
r''/r=-4\pi T_{ab}k^ak^b,\eqno(3)
$$ 
still holds since the Einstein equations do and $k^a$ is still null. 
One might even be able to get away with taking the form of $T_{ab}$ to be 
that of an isotropic, homogeneous perfect fluid, putting in the 
inhomogeneity spatial dependence by hand.  That is, write the RHS as 
$-4\pi\,(1+z)^2\rho(z)$, equivalent to the second part of equation (1).  

However, we cannot simply apply this sort of patch onto the FRW model.  Now 
the relation between affine 
parameter and redshift, $d\lambda/dz=(1+z)^{-2}H(z)$, is not the global 
FRW one (nor is it isotropic if the inhomogeneity field is not).  Thus we 
cannot directly compare to the FRW situation in terms of the observable, 
the redshift, because we have different 
dynamics; the redshift to the source will differ if the observer is in 
a (dynamically significant) inhomogeneity.  

Furthermore, if we try matching boundaries between an inhomogeneous and 
homogeneous region, we see the whole model breaks down: from the Friedmann 
equations a jump in density implies a discontinuity in the expansion rate 
$\dot a$ at the 
boundary and the two regions will (unphysically) lose contact.  The 
Robertson-Walker metric has broken down and we must instead use another 
that takes into account the impact of the inhomogeneities, 
e.g.~the Tolman-Bondi case.  This is what MBHE did and they 
found, not surprisingly, that no reduction to FRW behavior occurred.  
Different dynamics implies different light behavior regardless of 
similarities in the large scale volume averages. 

Similarly, note that questions of flux conservation become more complicated 
when the dynamics of the models differ; one cannot simply compare areas 
at some fixed redshift $z$.  
So if we hope to obtain an asymptotic agreement with FRW behavior then 
we must keep our inhomogeneities small enough that the dynamics is 
undisturbed.  In this case the DR model may be 
a reasonable approximation.  Let us examine further the role of different 
types of averaging. 

\bigskip 
\centerline{\bf 3. Averaging}
\medskip 
\centerline{\bf 3.1 Volume Averaging}
\medskip 
As we have just seen, in 
comparing two dynamically different universes there should be 
no requirement or expectation that they produce the same large scale 
light propagation behavior.  One example of this is the MBHE result. 
But another significant point of that result is 
the breakdown of volume averaging.  Smoothing over the inhomogeneity 
scale reduces the gross physical conditions and dynamics 
to FRW, but a difference in distance behavior remains.  Thus volume 
averaging is not sufficient to enforce equal distance-redshift relations, 
because the light also probes the small scale structure of the spacetime. 

However, MBHE take a highly nongeneric situation of spherical symmetry of 
the inhomogeneity around the observer.  But it is precisely the 
observer who is comparing distances and light propagation behaviors.  
The observer is indeed in a privileged position and one cannot carry out 
such Copernican averaging.  While volume averaging may be 
correct within an ergodic framework, this is 
fundamentally unobservable -- we observe at a single location.  

So such explicit volume averaging seems overly restrictive. 
To avoid this let us examine the distance-redshift 
relation in a more generic model.  
Consider the DR model, which possesses 
nonexplicit volume averaging, i.e.~every substantial volume is taken to 
look like FRW under coarse graining.  Imagine one case where the observer 
lives in a finite inhomogeneous region and another case where the observer 
lives in a FRW region but surrounded by an inhomogeneous shell at some 
distance.  

Within the DR model the beam equation for the light propagation (see eq.~2) 
depends 
only on the global expansion rate $H(z)$, deceleration parameter $q(z)$, 
density parameter $\Omega(z)$ (all of which are related), and the local 
inhomogeneity pattern through the density clumping parameter along the line 
of sight, $\alpha$.  We will sometimes find it useful to employ the 
generalization of DR to arbitrarily varying clumpiness $\alpha(z)$ 
(Linder 1988a).  
[Except under isotropy or angle averaging (see the next 
section), $\al(z)$ will really be a function of position on the sky as well: 
$\al(z; \vec\th, \vec\phi)$.  However, the DR model encompasses an implicit 
isotropy through its implicit volume averaging.  In any case, we do not 
need that: 
it suffices for us to consider 
a given line of sight with some $\al(z)$; another line of sight could have 
$\al'(z)$.  We do not assume the observer lives in any special 
position.] 

Linder (1988a) gives a closed form integral solution for 
the angular distance $r(z)$ in terms of $\al(z)$, the global parameters, 
and the FRW relative distance $r_{frw}(z',z)$.  Alternatively one can simply 
solve the beam equation (2), or even obtain an analytic solution if one 
considers discrete regions with different, constant $\alpha$ ($\alpha=1$ for 
FRW) and matches them at the boundaries.  

If one uses discrete regions one needs to match 
the distances and their first derivatives (equal to the local Hubble flow) 
at the boundary redshift (there is no ambiguity in the redshift 
since the global dynamics of the regions are the same).  
While this is fine for the analytic solution, it presents problems for 
the closed form solution.  The beam equation  
is a second order equation in the distance.  Despite matching 
the zeroth and first derivatives, the discontinuity in the density 
($\alpha$) at the boundary requires a discontinuity in the second 
derivative of the distance.  This invalidates the integral solution 
since then $\ddot r(z)$ or $\alpha(z)$ is undefined at the boundary.  Of 
course it 
is trivial to fix this by taking a rapid but smooth change in $\alpha$ 
between the desired regions.  

We take a toy model of a transition in $\al$ from $\al_1$ in the vicinity 
of the observer to $\al_2$ far from the observer along the line of sight. 
We calculate the distance-redshift relation using both a step in $\al$ at 
$z_*$, evaluating the distance by the analytic matching procedure, and a 
smoother transition in $\al$, 
$$ 
\al(z)=\al_1\,[1+e^{(z-z_*)/\Dl}]^{-1}+\al_2\,[1+e^{-(z-z_*)/\Dl}]^{-1},
\eqno(4)
$$ 
numerically solving the beam equation.  Here $\Dl$ is the transition width. 
As $\Dl\to0$ the two approaches agree. 

As a first case consider a sphere of inhomogeneity $\alpha=0$ surrounding the 
observer out to redshift $z_*$, with beyond $z_*$ the universe perfectly 
FRW ($\al=1$).  
(Again, the sphericity is not important, as we could just consider 
$\al(z)$ along some particular line of sight.)  
Even for $z_*$ small the distance to a source at $z>z_*$ is not found to 
approach the FRW distance even asymptotically as $z\gg z_*$, but maintains a 
constant 
fractional offset of order $z_*^2$.  While general inhomogeneous universes 
would require $z_*$ to be small in order to recover FRW dynamics (if those 
were used in the calculation, as here), this is not required in the DR 
model since its implicit volume averaging, i.e.~density compensation, 
automatically gives FRW dynamics.  
In any case, this has 
the important implication that even for a universe that agrees with FRW 
in both dynamics and as a volume average 
there can be a situation where no line of sight obeys the FRW relation. 

As the second case suppose that the observer lives in a FRW region but a 
small locality 
along a line of sight has inhomogeneity given by the clumpiness $\alpha$ 
of the DR model.  Here the observer is certainly not in a special position.  
If the clumpiness, say $\alpha=0$, extends from $z_1$ to $z_1+\Delta z$, 
then the solution for the angular distance to sources beyond the small 
patch of inhomogeneity again does not converge to the FRW result. 
Instead it asymptotically has a constant fractional offset of order 
$\Delta z/(1+z_1)$. 
That it does not agree with FRW does not depend on each line of sight 
having the same 
inhomogeneity location $z_1$, i.e.~the observer being at the center of 
a spherical shell of inhomogeneity, so the observer truly is not singled 
out in any Copernican sense.  So long as every line of sight 
had a clumpy region somewhere along it, the (anisotropic) distance-redshift 
relation will nowhere be FRW. 

These cases show that despite universes looking like FRW under volume 
averaging, their distance-redshift relations may not be FRW. 

\bigskip 
\centerline{\bf 3.2 Angular Averaging} 
\medskip 
Taken at face value, the DR model seems to imply that since each line of 
sight has a distance greater than FRW at the same redshift, even an angle 
averaging procedure cannot restore the FRW behavior.  However, as mentioned 
in Section 1, the DR approach ignores those lines of sight passing near the 
inhomogeneity clumps.  Once those are included, then the full sky average 
distance relation does agree with FRW, and flux conservation holds 
since the universes are dynamically equivalent.  [Note that 
flux conservation requires $<r_a^{-2}>=r_{frw}^{-2}$; there is no 
requirement that $<r_a>=r_{frw}$, though of course if $r_a=r_{frw}$ for 
each line of sight then flux conservation is automatic.] 

Even without a full sky average the DR distance relation approaches FRW 
upon angle averaging over a sufficiently large angular scale, i.e.~a wide 
enough observation beam.  One criterion for this scale is given 
by Linder (1998).  Here we briefly mention two further points regarding the 
transition of the DR relation to FRW.  

In that paper, the angular averaging scale was equated to an angular scale 
where the beam had a certain probability to include a clump and hence feel 
the full density, smooth plus clumpy.  This was done under Poisson 
statistics.  Here we note that if the clumps cluster, as galaxies for 
example, then the excess mass density in an angular region is proportional 
to the angular two point mass correlation function $w(\th)$, evaluated at 
the source depth.  This helps determine the angular transition scale to FRW 
behavior.  Secondly, note that the shear contribution to the effective 
smoothness $\al$, equation (12) in that paper, dies off as $(1+z)^{-5}$ 
for constant shear and as $(1+z)^{-7}$ for shear in the linear density 
perturbation regime. 

\bigskip 
\centerline{\bf 3.3 Line of Sight Averaging} 
\medskip 
As discussed in the previous section, while the DR model does implicitly 
volume average to a FRW universe, and does approach the FRW distance relation 
for sufficiently large angular averages, it does not along a given line of 
sight.  Let us ask if there is any way to restore FRW behavior for every 
line of sight (short of taking a purely FRW universe). 

From the closed form expression of Linder (1988a) for the deviation of 
the angular distance from the FRW behavior, 
$$
r(z)-r_{frw}(z)=(3/2)\omo\int_0^z ds\,(1+s)(1+\omo s)^{-1/2}[1-\al(s)] 
r(s)\,r_{frw}(s,z),\eqno(5)
$$
in a dust universe, one sees that the integrand must oscillate in such a 
way as to cause the RHS to vanish.  For a localized inhomogeneity with 
density smoothness parameter $\al$, this can be done by requiring a 
neighboring inhomogeneity to compensate for the density deviation from FRW 
suffered by the beam.  This condition on the two regions becomes, for small 
inhomogeneity extents $\Dl z$, 
$$
(1-\al_1)\,\Dl z_1=-(1-\al_2)\,\Dl z_2.\eqno(6)
$$
We will call this density compensation or line of sight averaging. 

This is not a particularly elegant or transparent condition; it does not 
correspond to a simple physical situation like requiring the same projected 
mass along the line of sight as the FRW model.  In fact, that property of 
having the same projected mass in no way guarantees that the light 
propagation behavior will be similar -- the {\it run} of density is what is 
important.  Furthermore, there is no particular 
reason to believe that a realistic inhomogeneous universe would naturally 
include density compensation. 

The DR model does not generically satisfy this condition for the density 
along the line of sight, but we can evaluate a model with 
compensated DR inhomogeneities, i.e.~ones arranged to obey equation (6). 
Along every line of sight we allow there to be a DR inhomogeneity 
region where $\alpha$ is less than one. 
We then compensate for this ``lost mass'' by taking 
an adjacent region to have higher than global density, arranging the 
size (extent in redshift) of this region according to equation (6) to make 
up exactly for the mass 
deficit from FRW experienced by the light ray in the first region.  

There are a couple of technical points to clarify before proceding to 
the results.  The first region, with $\alpha<1$, does not actually lack 
any mass; the DR model says the inhomogeneous region has the same 
mass and density in the volume averaged sense as the equivalent FRW 
volume.  So there seems no a priori reason why such a compensation would 
be required for the light ray -- {\it within the framework of volume  
averaging being a legitimate treatment for evaluating light propagation 
behavior}.  Nevertheless, we will see that compensation is key, so the 
volume averaging procedure is lacking.   

Secondly, the compensating 
region will have $\alpha>1$, that is the light ray experiences a density 
above the FRW average.  Mathematically, it is fine to treat this by 
means of the usual DR clump interpretation but with clumps of negative 
mass.  This is physically a bit unsettling but there are two ways around 
it.  One, one can make the clumps small but not pointlike, as the usual 
DR model uses, and then they can simply have density sufficiently below 
the FRW value that this mass deficit offsets the mass excess from the 
smoothly distributed above average density matter.  Second, one can 
ignore the interpretation of the density deviations as being in clumps 
plus a smooth component and simply treat $\alpha\Omega$ in the beam equation 
as giving the actual density along the light path in this inhomogeneity 
region.  Then there is 
no problem having $\alpha$ less than or greater than one.  

[Note that in 
the usual solution to the distance-redshift relation the expression 
breaks down for $\alpha>25/24$.  We are free to take our compensating 
region to have $1<\alpha<25/24$ with a correspondingly greater redshift 
extent, but we will find that our final solution is independent of the 
$\alpha$ adopted.  So we need not specify it or worry about the upper limit, 
in line with the second interpretation above.]

We calculate two situations: 1) the observer is in a region with 
$\alpha=0$, extending out to $\Delta z$, which is then compensated by an 
adjoining region according to 
the above prescription; 2) the observer is in a FRW region but at some $z_1$ 
the light ray passes through an inhomogeneity with $\alpha=0$ and extent 
$\Delta z$, which is compensated by an adjoining region according to 
the above prescription.  In both cases the universe is taken to be FRW 
outside of the localized inhomogeneities.  These cases are the analogs of the 
uncompensated ones calculated above in section 3.1.  

With the compensations, however, we find that the distance-redshift 
behavior beyond the inhomogeneity agrees exactly with the FRW result. 
Note that no assumption was made of spherical symmetry or isotropy; 
the compensation prescription works independently for each line of 
sight, as designed.  As mentioned in section 3.2, flux conservation is then 
automatic.  In this light, one can consider the large angle averaging 
approach to FRW of Linder (1998) as another method of ensuring compensation: 
widening the line of sight, i.e.~the beam, allows the effective density 
smoothness to 
approach the FRW value naturally within the DR model, without the need 
for a secondary inhomogeneity region. 
In some sense they are complementary; averaging {\it along} 
the line of sight through compensation, or averaging {\it transverse} to 
the line of sight by beam widening. 

For an alternate view of line of sight averaging, we can use the 
postNewtonian perturbed approach of Jacobs, Linder, \& Wagoner (1993).  
Here, light propagation can be calculated from a metric realistically 
describing an inhomogeneous universe so long as the gravitational potential 
$\phi$ of the inhomogeneities is small compared to unity and its first 
derivatives small compared to the Hubble scale $H_0^{-1}$ (this condition 
also ensures that the dynamics are driven by the background, FRW model).  
Given the metric, one can derive the angular distance deviation to be 
$$
r(z)-r_{frw}(z)=-H_0^{-2}\int_1^{1+z} du\,u^{-1/2}\,(\nabla^2\phi)\,r(u) 
r_{frw}(u,1+z),\eqno(7)
$$
in a flat, dust background.  The right hand side, giving the deviation 
from FRW distance behavior, arises from the perturbed Ricci tensor.  

The condition for approaching FRW is that the 
integration over the path length must reduce the possibly large (compared 
to background) but fluctuating effective density deviation, 
$\nabla^2\phi$.  This would occur naturally unless the density 
fluctuations were arranged coherently over the path length.  Otherwise, 
the RHS would be expected to be of order $\nabla\phi$, which is small by 
original assumption.  That is, one has sufficient effective mass compensation 
(oscillation), as in the above line of sight averaging procedure, that the 
distance deviation becomes small.  So in a realistically inhomogeneous 
universe one expects the distance-redshift 
relation to approach FRW for path lengths much larger than the inhomogeneity 
coherence scale. 

\bigskip 
\centerline{\bf 4. Conclusion} 
\medskip 
The results of the three categories of models for an inhomogeneous universe 
can be summed up concisely: 
\smallskip 
$\bullet$\quad A model that is FRW in a volume averaged sense but with 
different dynamics (e.g.~{\bf Tolman-Bondi}) does not yield FRW results for 
light propagation. 
\smallskip 
$\bullet$\quad A model that is FRW in a volume averaged sense and also agrees 
dynamically (e.g.~{\bf Dyer-Roeder}) but does not match in a line of sight 
averaged sense does not generally yield FRW results for light propagation. 
However, it can if angle averaging provides effective line of sight 
density compensation. 
\smallskip 
$\bullet$\quad A model that is FRW in a line of sight averaged sense (same 
local mass) and 
agrees dynamically ({\bf compensated Dyer-Roeder}) will yield FRW results 
for light propagation.  This seems to be a rather ad hoc model though. 
Depth averaging, however, where the path length exceeds the inhomogeneity 
coherence length, can create effective compensations naturally. 
\medskip 

Because light propagation probes the 
spacetime along the path on all scales, a coarse graining by volume averaging 
over a locally inhomogeneous universe will not necessarily yield the light 
propagation of that average spacetime.  That is, a large scale volume 
averaging that 
smooths the universe to FRW, while maybe sufficient for dynamical 
(e.g.~expansion and growth of perturbation) calculations, is not appropriate 
for recovering FRW light propagation behavior.  One must adopt a line of 
sight averaging that brings the path conditions to FRW in order to match 
the FRW result. 

This can be achieved through 1) ensuring that density 
perturbations along the line of sight are compensated, 2) smoothing the 
perturbations by using beams sufficiently 
wide to sample beyond the extent of the inhomogeneity (essentially the 
wide beam just redefines what is meant by line of sight), 
or 3) {\it local} volume averaging where the inhomogeneities can be 
smoothed on scales small enough to retain the background dynamics and 
provide effective mass compensation within the beam.  Again, the local mass 
agreement with 
the background is required because light probes the local, not just 
average, gravitational field.  

As one example, for the Tolman-Bondi model of MBHE none of these hold: 
the dynamics is not FRW and the effective coherence scale is infinite 
due to the special position of the observer.  Generally, 
if the physical conditions of an observation, in angular width and path 
characteristics, do not match one of the above situations then one should not 
expect to be able to legitimately use a FRW distance-redshift relation to 
interpret that observation. 

\medskip 
This work was supported by NASA grants NAG5-3525, NAG5-3922, and NAG5-4064. 
\vfill\break 
\centerline{\bf REFERENCES} 
\parindent=0in 
\bigskip 
Bildhauer, S.~\& Futamase, T.~1991, MNRAS, 249, 126

Buchert, T.~1997, in Proc.~2nd SFB Workshop on Astro-particle Physics, 
\vskip-2pt 
\qquad ed.~R.~Bender et al., p.~71 (astro-ph/9706214) 

Buchert,T.~\& Ehlers, J.~1997, A\&A, 320, 1

Dyer, C.C.~\& Roeder, R.C.~1972, ApJ, 174, L115 

Dyer, C.C.~\& Roeder, R.C.~1972, ApJ, 180, L31 

Fukushige, T.~\& Makino, J.~1994, ApJ, 436, L111 

Jacobs, M.W., Linder, E.V., \& Wagoner, R.V.~1993, PRD, 48, 4623 

Linder, E.V.~1988a, A\&A, 206, 190 

Linder, E.V.~1988b, A\&A, 206, 199

Linder, E.V.~1998, ApJ, 497, in press (astro-ph/9707349) 

Mustapha, N., Bassett, B.A., Hellaby, C., \& Ellis, G.F.R.~1997, 
\vskip-2pt 
\qquad submitted to Class.~Q.~Grav., gr-qc/9708043 

Watanabe, K.~\& Tomita, K.~1990, ApJ, 355, 1
\bye